\documentclass[aps,prl,showpacs,twocolumn,nofootinbib,floatfix,superscriptaddress]{revtex4}
\usepackage{amsmath,graphicx,epsfig,amssymb,subfigure,times}
\usepackage[usenames]{color}

\newcommand{\h}[2][ ]{\hat{#2}^{\vphantom{\dag} #1}}
\newcommand{\hd}[2][ ]{\hat{#2}^{\dag #1}}

\begin{document}

\title{Detection of continuous variable entanglement without coherent local oscillators}

\author{Andrew~J.~Ferris}
\author{Murray~K.~Olsen}
\affiliation{The University of Queensland, School of Physical Sciences, ARC Centre of Excellence for Quantum-Atom Optics, Qld 4072, Australia}

\author{Eric~G.~Cavalcanti}
\affiliation{The University of Queensland, School of Physical Sciences, ARC Centre of Excellence for Quantum-Atom Optics, Qld 4072, Australia}
\affiliation{Centre for Quantum Dynamics, Griffith University, Qld 4111, Australia}

\author{Matthew~J.~Davis}
\affiliation{The University of Queensland, School of Physical Sciences, ARC Centre of Excellence for Quantum-Atom Optics, Qld 4072, Australia}

\date{\today}

\begin{abstract}
We propose three criteria for identifying continuous variable entanglement between two many-particle systems with no restrictions on the quantum state of the local oscillators used in the measurements. Mistakenly asserting a coherent state for the local oscillator can lead to incorrectly identifying the presence of entanglement.  We demonstrate this in simulations with 100 particles, and also find that large number fluctuations do not prevent the observation of entanglement. Our results are important for quantum information experiments with realistic Bose-Einstein condensates or in optics with arbitrary photon states.
\end{abstract}

\pacs{03.67.Mn,03.75.Gg,03.65.Ud,42.50.Dv}

\maketitle

%\section{Introduction}
The study of the quantum properties of matter-waves is a rapidly developing field known as quantum atom optics~\cite{Rolston2002a,Molmer2003a}.  Already several experiments have observed non-classical effects in ultra-cold gases, including the Hanbury Brown-Twiss effect for bosons~\cite{Schellekens2005a}, anti-bunching for fermions~\cite{Vassen2007a}, sub-Poissonian number fluctuations~\cite{Chuu2005a}, and density correlations from molecular dissociation~\cite{Greiner2005a} and in the Mott-insulator regime in an optical lattice~\cite{Folling2005a}. Although impressive achievements, the experimental techniques utilized in these observations are insufficient to detect quantum squeezing or entanglement. The demonstration of entanglement ---  which Schr\"{o}dinger described as being the central mystery of quantum mechanics~\cite{Schroedinger1935a} --- will be an important step towards quantum information applications of ultra-cold atomic systems.

In quantum optical systems continuous variable (CV) entanglement can be demonstrated experimentally by measuring certain correlation functions of electromagnetic field quadratures, and finding that these violate an inequality for separability, \cite{Duan2000a,Simon2000a}, or an inequality \cite{Reid1989a} for demonstrating the Einstein-Podolsky-Rosen (EPR) paradox~\cite{Einstein1935a}.  For a number of systems, these quadratures can only be determined using homodyne or heterodyne measurement techniques, and require a coherent state local oscillator --- a highly occupied mode of the electromagnetic field that is a good approximation to the output of many lasers~\cite{QuantumNoise}. Such measurements have led to the observation of optical squeezing~\cite{Shlusher1985a,Zhang2001a} and the EPR paradox with photons~\cite{Ou1992a}, and have been used to perform quantum state tomography~\cite{Smithey1993a} and continuous-variable teleportation~\cite{Bowen2003a}.

In principle squeezed or entangled atomic fields can be generated from atomic four-wave-mixing~\cite{Hilligsoe2005a,Campbell2006a,Olsen2006a}, molecular disassociation~\cite{Kheruntsyan2005a}, or by mapping photon statistics onto atoms~\cite{Haine2005a}.  Entanglement generated in these situations can potentially be used as a resource for quantum information~\cite{Braunstein2005a}.  However, to unequivocally demonstrate entanglement between matter-waves it will be necessary to perform measurements sensitive to the relative phase of atomic wave packets. In principle it is possible to measure matter-wave quadratures in direct analogy to the optical case using atomic measurements with a suitable local oscillator (phase reference)~\cite{deOliveira2003a,daCunha2007a}, of which the matter-wave equivalent is a Bose-Einstein condensate.  However these are typically not large, with a maximum of $10^8$ particles~\cite{vanDerStam2007a}, and the phase stability is compromised compared to a laser, as atomic interactions result in a mode shape and energy that depend on the particle number. Unfortunately, the usual correspondence between homodyne measurements and the field quadratures that is ubiquitous in quantum optics may be lost if the local oscillator is neither large nor coherent.

In this paper we show how to detect CV entanglement without a coherent state as a phase reference. We introduce three new entanglement criteria  based on homodyne measurement that require no assumptions about the quantum state of the local oscillator.  These allow for the demonstration of inseparability and the EPR paradox with measurements that are feasible with current or foreseeable technologies with ultra-cold atoms.  We give a numerical demonstration of the application of these criteria in an experiment with 100 particles.

\begin{figure}[t]
\begin{centering}
\includegraphics[width=7cm]{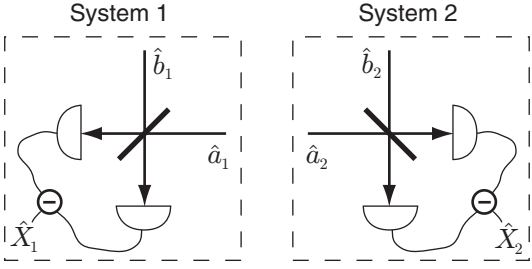}
\caption{Homodyne measurements of the two systems.  Signal modes $1$ and $2$ are mixed with local oscillators; the resulting number differences are a measurement of the quadratures $\h{X}_1$ and $\h{X}_2$.}
\label{fig:system}
\end{centering}
\end{figure}

We begin by reviewing homodyne measurement methods. The two quadrature operators for the mode with annihilation operator $\h{a}$ are canonically defined to be
\begin{equation} \label{quad_c}
  \h{X}_{\mathrm{c}} = \h{a} + \hd{a}, \hspace{0.5cm}\mathrm{and}\hspace{0.5cm} \h{Y}_{\mathrm{c}} = i\left(\h{a} - \hd{a}\right).
\end{equation}
These operators do not conserve boson number and are not detectable directly by intensity measurements. In optics, homodyne and heterodyne measurements allow access to these observables by using a beam splitter to interfere the signal with a local oscillator with well-defined phase and amplitude followed by intensity measurement(s) {as illustrated in Fig.~\ref{fig:system}}. Using the Heisenberg picture, if $\h{a} \equiv \h{a}_{\mathrm{in}}$ and $\h{b} \equiv \h{b}_{\mathrm{in}}$ describe the signal and local oscillator before the beam splitting, then output modes $\h{a}_{\mathrm{out}}$ and $\h{b}_{\mathrm{out}}$ are given by
\begin{equation} \label{beamsplit}
 \h{a}_{\mathrm{out}} = t \h{a}_{\mathrm{in}} + r \h{b}_{\mathrm{in}}, \hspace{0.5cm}\mathrm{and} \hspace{0.5cm} \h{b}_{\mathrm{out}} = r^{\ast} \h{a}_{\mathrm{in}} - t^{\ast} \h{b}_{\mathrm{in}},
\end{equation}
where $|t|^2 + |r|^2 = 1$. Here we focus on balanced homodyning using a 50-50 beam splitter with $t = r = 1/\sqrt{2}$.  The final step of a balanced homodyne measurement is to measure the difference in the number of particles exiting the beam splitter ports $\h{a}_{\mathrm{out}}$ and $\h{b}_{\mathrm{out}}$. The measured quadrature is rescaled by the size of the local oscillator, according to
\begin{equation} \label{quad}
  \h{X}_{\mathrm{m}} = \frac{\hd{a}_{\mathrm{out}} \h{a}_{\mathrm{out}} -
\hd{b}_{\mathrm{out}}\h{b}_{\mathrm{out}}}{\langle \hd{b}_{\mathrm{in}} \h{b}_{\mathrm{in}}\rangle^{1/2}} =
\frac{\hd{a}\h{b} + \h{a}\hd{b}}{\langle \hd{b} \h{b}\rangle^{1/2}}.
\end{equation}
If the local oscillator is a coherent state $|\beta\rangle$, where $\beta$ is real and positive,
and the coherent state is large $\langle\hd{b}\h{b}\rangle = |\beta|^2 \gg \langle \hd{a}\h{a} \rangle$,
then the difference between the two quadrature operators $\h{X}_{\mathrm{c}}$ and $\h{X}_{\mathrm{m}}$ will be negligible.  For a signal that is comparable in size to a coherent local oscillator $|\beta|^2 \sim \langle \hd{a}\h{a} \rangle$, the moments of the canonical quadrature observables can be determined from the measured observables using the result \cite{Zhang2001a,QuantumNoise}
\begin{equation}
   \langle \h{X}_{\mathrm{m}} \rangle = \langle \h{X}_{\mathrm{c}} \rangle, \quad\quad \langle \h[2]{X}_{\mathrm{m}} \rangle = \langle \h[2]{X}_{\mathrm{c}} \rangle +
{\langle \hd{a}\h{a} \rangle}/{\langle \hd{b}\h{b} \rangle}.
%  \frac{\langle \hd{a}\h{a} \rangle}{\langle \hd{b}\h{b} \rangle}.
\label{moments}
%  \Bigl\langle \bigl(\h{X}_{\mathrm{m}} - \h{X}_\mathrm{c}\bigr)^2 \Bigr\rangle = \frac{\langle \hd{a}\h{a} \rangle}{\langle \hd{b}\h{b} \rangle}.
\end{equation}
However, for local oscillators that are not coherent states the difference between
$\h{X}_{\mathrm{m}}$ and $\h{X}_{\mathrm{c}}$ cannot be neglected.

For matter-waves the equivalent of the beam splitter operation is performed by Bragg, RF, or Raman pulses~\cite{Rolston2002a} to interfere atoms with different momenta and/or internal states.  However, the best BEC phase references will typically have less than $10^7$ atoms.
With the additional effects of atomic interactions and finite temperatures, it can be expected that typical BECs in the lab will have greater phase and/or amplitude fluctuations than the coherent output of a laser.  Therefore we must take full account of the difference between the canonical and measured quadrature variables when detecting entanglement in condensate experiments.

\textit{Separability criterion:} The observation of CV entanglement requires the simultaneous measurement of two quadratures, and so balanced homodyning requires two phase references. We follow the general method of Duan \emph{et al.}~\cite{Duan2000a}, Simon~\cite{Simon2000a} and Hofmann and Takeuchi~\cite{Hofmann2003a} to derive a separability criterion for the two systems of interest, depicted in Fig.~\ref{fig:system}. System~1 consists of the signal $\h{a}_1$ and phase reference $\h{b}_1$, and System~2 has signal $\h{a}_2$ and phase reference $\h{b}_2$. Measurements are made of the (number-conserving) quadratures
\begin{equation} \label{measured}
  \h{X}_j = \frac{\h{a}_j \hd{b}_j + \hd{a}_j \h{b}_j}{\langle \hd{b}_j
\h{b}_j\rangle^{1/2}}, \hspace{1cm} %\\
  \h{Y}_j = i\frac{\h{a}_j \hd{b}_j - \hd{a}_j
\h{b}_j}{\langle \hd{b}_j \h{b}_j\rangle^{1/2}},
\end{equation}
with commutator
%\begin{equation}
$
  \left[ \h{X}_j, \h{Y}_k \right] = - 2 i \delta_{j,k} ({ \hd{b}_j \h{b}_j -
\hd{a}_j \h{a}_j})/{\langle \hd{b}_j \h{b}_j \rangle}.
$
%\end{equation}

It has been shown~\cite{Raymer2003a} that for any local observables on the two systems, $\h{A}_j$ and $\h{B}_j$, with commutator $\bigl[\h{A}_j,\h{B}_j\bigr] = i\h{C}_j$, all separable states obey
\begin{equation}
  \mathrm{Var}\bigl[\h{A}_1 + \h{A}_2\bigr] + \mathrm{Var}\bigl[\h{B}_1 + \h{B}_2\bigr] \ge |\langle \h{C}_1 \rangle| + |\langle \h{C}_2 \rangle |,
\end{equation}
where $\mathrm{Var}[\h{A}]$ is the variance of the observable corresponding to the operator $\h{A}$. For our system, it follows that the two systems are entangled when the inequality
\begin{multline} \label{Duan2}
  \mathrm{Var}\bigl[\h{X}_1 \pm \h{X}_2\bigr] + \mathrm{Var}\bigl[\h{Y}_1 \mp \h{Y}_2\bigr]
  \ge \\ \;2\left| 1 - {\langle\hd{a}_1
\h{a}_1\rangle}/{\langle\hd{b}_1\h{b}_1\rangle} \right|
  + 2 \left| 1 - {\langle\hd{a}_2
\h{a}_2\rangle}/{\langle\hd{b}_2\h{b}_2\rangle} \right|.
\end{multline}
is violated.   For semi-classical local-oscillators the right hand side (RHS) of Eq.~(\ref{Duan2}) approaches four, as originally derived by Duan \emph{et al.}~\cite{Duan2000a} and Simon~\cite{Simon2000a}. Note that this result is not the same as applying the coherent state corrections [Eq.~(\ref{moments})] to the derivation of Duan \emph{et al.}~\cite{Duan2000a} and Simon~\cite{Simon2000a} --- for further discussion see the Appendix.

%The lower bound of the variances in Eq.~(\ref{Duan2}) is not always zero. The Heisenberg uncertainty principle gives
%\begin{multline}
%\mathrm{Var}\bigl[\h{X}_1 \pm \h{X}_2\bigr] + \mathrm{Var}\bigl[\h{Y}_1 \mp
%\h{Y}_2\bigr] \ge
% 2 \left| \frac{\langle\hd{a}_1 \h{a}_1\rangle}{\langle\hd{b}_1 \h{b}_1\rangle}
%- \frac{\langle\hd{a}_2 \h{a}_2\rangle}{\langle\hd{b}_2 \h{b}_2\rangle} \right|,
%\end{multline}
%which for some states is equal to the bound in Eq. (\ref{Duan2}). However, this lower bound approaches zero for large local oscillators, or when the ratio of the signal to local oscillator size is equal in each system.

\textit{EPR criteria:} The value of a quadrature of System~2  may be estimated from
measurements of System~1 due to the existence of correlations between them. For separable states, the accuracy of this
estimation will be restricted by the Heisenberg uncertainty limit.
However, entangled systems may allow one to make predictions
seemingly better than this fundamental limit, a fact related to the
Einstein-Podolsky-Rosen paradox~\cite{Einstein1935a} by Reid~\cite{Reid1989a}.

We denote the inferred value of $\h{X}_2$ by the measurement result
of $\h{X}_1$ as $\h[\mathrm{inf}]{X}_2 = f(\h{X}_1)$, and similarly
$\h[\mathrm{inf}]{Y}_2 = g(\h{Y}_1)$, with the error of these estimates  given by
$\Delta_{\mathrm{inf}}^2\bigl[\h{X}_2\bigr] = \langle (\h{X}_2
-\h[\mathrm{inf}]{X}_2)^2\rangle$ and
$\Delta_{\mathrm{inf}}^2\bigl[\h{Y}_2\bigr] = \langle
(\h{Y}_2-\h[\mathrm{inf}]{Y}_2)^2\rangle$, respectively.
The EPR paradox is demonstrated when the inferred predictions have better accuracy than allowed by
the Heisenberg uncertainty principle~\cite{Reid1989a}.  This occurs when the inequality
\begin{equation} \label{eq:EPR1}
  \Delta_{\mathrm{inf}}^2\bigl[\h{X}_2\bigr] \,
  \Delta_{\mathrm{inf}}^2\bigl[\h{Y}_2\bigr]
  \ge \left( 1 - {\langle \hd{a}_2 \h{a}_2
\rangle}/{\langle\hd{b}_2\h{b}_2\rangle} \right)^2,
\end{equation}
is violated. Although the optimal inferred quadrature values could be in principle any function of the measured values, for simplicity we
restrict ourselves to linear functions, i.e. $\h[\mathrm{inf}]{X}_2 = a + b\h{X}_1$ and $\h[\mathrm{inf}]{Y}_2 = c + d\h{Y}_1$. The simplest guess can be motivated by the inseparability criterion above; if the variances in Eq.~(\ref{Duan2}) were vanishingly small then we would infer that $\h[\mathrm{inf}]{X}_2 = \mp\h{X}_1 \pm \langle\h{X}_1\rangle + \langle\h{X}_2\rangle$ and $\h[\mathrm{inf}]{Y}_2 = \pm\h{Y}_1 \mp \langle\h{Y}_1\rangle + \langle\h{Y}_2\rangle$. This results in inference errors
\begin{gather}
  \Delta^2_\mathrm{inf}\bigl[\h{X}_2\bigr] = \mathrm{Var}\bigl[\h{X}_1\pm \h{X}_2\bigr],
  \nonumber\\
  \Delta^2_\mathrm{inf}\bigl[\h{Y}_2\bigr] = \mathrm{Var}\bigl[\h{Y}_1\mp \h{Y}_2\bigr].\label{DuanEPR2}%\\
\end{gather}
Together with Eq. (\ref{eq:EPR1}), this implies
\begin{equation} \label{DuanEPR}
  \mathrm{Var}\bigl[\h{X}_1 \pm \h{X}_2\bigr] + \mathrm{Var}\bigl[\h{Y}_1 \mp \h{Y}_2\bigr]
  \ge 2\left| 1 - \frac{\langle\hd{a}_2 \h{a}_2\rangle}{\langle\hd{b}_2\h{b}_2\rangle} \right|,
\end{equation}
This is the same result as for the separability criterion above, Eq.~(\ref{DuanEPR}), except the RHS is twice as small (typically).  Thus it is more difficult to demonstrate the EPR paradox than inseparability, but it proves the stronger result that either local causality is violated or quantum mechanics provides an incomplete description of System~2 (for a discussion of this hierarchy see~\cite{Wiseman2007a,Jones2007a}). The RHS of Eq.~(\ref{DuanEPR}) becomes two for semi-classical local operators which agrees with the work of Reid \emph{et al}.~\cite{ReidReview2008}. This inequality is a special case of a more general EPR criterion for arbitrary observables~\cite{CavalcantiUnpublished2008}.

It is possible to derive a second EPR criterion that is violated for a greater range of states than Eq.~(\ref{DuanEPR}). The optimal linear inference of $\h{X}_2$ based on the measurement of $\h{X}_1$ is~\cite{Reid1989a}
\begin{equation}
  \h[\mathrm{inf}]{X}_2 = \langle \h{X}_2 \rangle +
\frac{\mathrm{Var}\bigl[\h{X}_1,\h{X}_2\bigr]}{\mathrm{Var}\bigl[\h{X}
_1\bigr]} \left( \h{X}_1 - \langle \h{X}_1 \rangle \right),
\end{equation}
where the covariance $\mathrm{Var}\bigl[\h{A},\h{B}\bigr] = \langle \h{A} \h{B} \rangle - \langle \h{A} \rangle\langle \h{B} \rangle$. In separate measurements we can estimate $\h{Y}_2$ after measuring $\h{Y}_1$ in the same way.
%\begin{equation}
%  \h[\mathrm{inf}]{Y}_2 = \langle \h{Y}_2 \rangle + \frac{\mathrm{Var}\bigl[\h{Y}_1,\h{Y}_2\bigr]}{\mathrm{Var}\bigl[\h{Y}_1\bigr]} \left( \h{Y}_1 - \langle \h{Y_1} \rangle \right).
%\end{equation}
The inference errors from this method are
\begin{eqnarray}
  \Delta^2_\mathrm{inf}\bigl[\h{X}_2\bigr] &=& \mathrm{Var}\bigl[ \h{X}_2 \bigr] -
  {\mathrm{Var}\bigl[\h{X}_1,\h{X}_2\bigr]^2}/{\mathrm{Var}\bigl[\h{X}_1\bigr]},\nonumber\\
  \Delta^2_\mathrm{inf}\bigl[\h{Y}_2\bigr] &=& \mathrm{Var}\bigl[ \h{Y}_2 \bigr] - {\mathrm{Var}\bigl[\h{Y}_1,\h{Y}_2\bigr]^2}/{\mathrm{Var}\bigl[\h{Y}_1\bigr]},\label{eq:EPR}%\\
\end{eqnarray}
which are less than or equal to the comparable results in Eqs.~(\ref{DuanEPR2}). Substituting Eqs.~(\ref{eq:EPR}) directly into Eq.~(\ref{eq:EPR1}) results in an EPR criterion that is superior to Eq.~(\ref{DuanEPR}). This result agrees with that of Reid~\cite{Reid1989a} in the limit that the phase-reference is large and the RHS of the inequality is unity.

\textit{Numerical demonstration:}
We illustrate the use of these criteria in degenerate four-wave mixing with atoms~\cite{Hilligsoe2005a,Campbell2006a}, where condensate atoms in mode 0 collide and populate modes 1 and 2 respectively.  The Hamiltonian is
\begin{equation}
  \h{H} = i\chi(\h[2]{a}_0 \hd{a}_1 \hd{a}_2 - \hd[2]{a}_0 \h{a}_1 \h{a}_2 ),
\end{equation}
where $\h{a}_j$ is the annihilation operator for mode $j$ and $\chi$ represents the strength of the coupling. This model can be realized with a moving condensate in an optical lattice, and the outgoing modes 1 and 2 are predicted to be entangled~\cite{Olsen2006a}.

\begin{figure}[t]
\begin{centering}
\includegraphics[width=8.5cm]{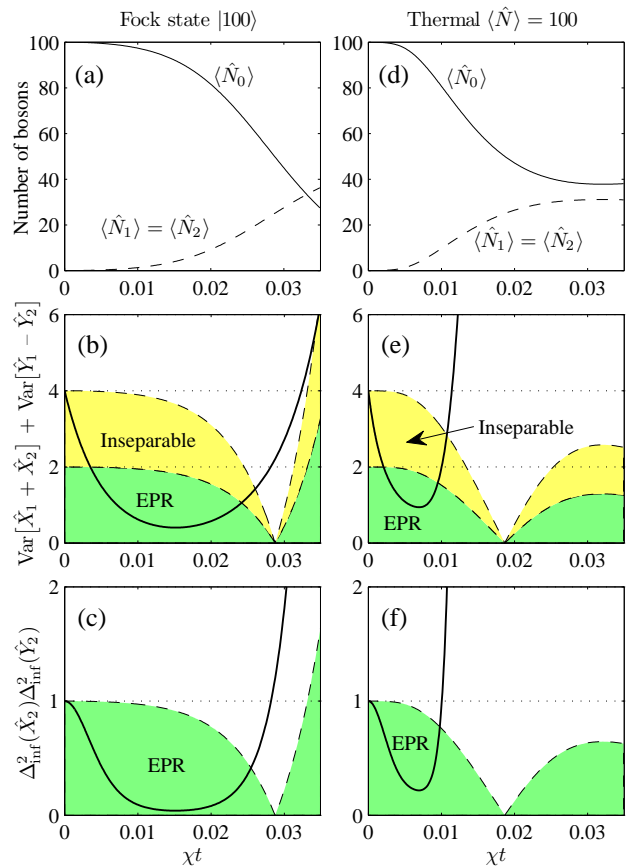}
\caption{(Color online) The population dynamics and entanglement criteria for the case of (a--c) exactly 100 particles and (d--f) a thermal distribution with a mean of 100 particles. In (a,d) we see that particles are transferred from the condensate to modes $1$ and $2$. In (b,c,e,f) the entanglement inequalities are violated when the solid line enters the shaded areas enclosed by the dashed lines.as described in the text.  The horizontal dotted lines indicate the inequality violation boundary for the canonical quadratures \protect{\cite{Duan2000a,Simon2000a,Reid1989a}}.} \label{fig:results}
\end{centering}
\end{figure}

Firstly, we consider an initial state with $N$ bosons in mode 0 and vacuum in 1 and 2 (i.e. $|\Psi(0)\rangle = |N\rangle|0\rangle|0\rangle$). The Hamiltonian then ensures the state can be written in the form
\begin{equation}
  |\Psi(t)\rangle = \sum_{m} c_m(t) |N-2m\rangle|m\rangle|m\rangle,
\end{equation}
at all times. The Hamiltonian in this basis can be numerically diagonalized (for $N \lesssim 10^4$), providing a solution for the exact dynamics of the system. Fig.~\ref{fig:results}(a) shows the number of atoms in each mode as a function of time for $N=100$.

To measure the quadratures of the signal modes (1 and 2) we require a phase reference for each. We can use the remaining condensate in mode 0 as our phase reference, but must first divide it in two with a beam-splitter according to
\begin{equation}
  \h{b}_1 = {(\h{a}_0 + \h{a}_3)}/{\sqrt{2}}, \hspace{1cm}
  \h{b}_2 = {(\h{a}_0 - \h{a}_3)}/{\sqrt{2}}.
\end{equation}
where the auxiliary mode 3 is initially in the vacuum state. Remember that these local oscillators are \emph{not} coherent states. We then perform the quadrature measurements as expressed by Eq.~(\ref{measured}), and compare the results with the separability criterion, Eq.~(\ref{Duan2}), and the EPR criteria, Eq.~(\ref{DuanEPR}) and Eqs.~(\ref{eq:EPR1},\ref{eq:EPR}). The results are plotted in Figures~\ref{fig:results}(b,c) where we can see that all the inequalities are violated at some stage in this experiment. After some time the number of particles in the signal beams grow larger than the phase references, and this measurement scheme is no longer optimal to detect the entanglement.

The weaker EPR criterion, Eq.~(\ref{DuanEPR}) in Fig.~\ref{fig:results}(b), shows a smaller region of $\chi t$ where violation occurs than the stronger version, Eq.~(\ref{eq:EPR}) in Fig.~\ref{fig:results}(c), as expected. The results in Figs.~\ref{fig:results}(b,c) identify times  where it is necessary to fully treat the quantum nature of the quadrature measurement. Performing these measurements and then using the criteria derived by Duan \emph{et al.}~\cite{Duan2000a}, Simon~\cite{Simon2000a} and Reid~\cite{Reid1989a} would lead to falsely identifying entanglement when the size of the phase references and signal beams become comparable.

A particularly interesting result is that the inseparability criterion [Eq.~(\ref{Duan2})] identifies entanglement in a small region [the top-right corner of Fig.~\ref{fig:results}~(b)] when the local oscillator is actually smaller than the signal modes and has sufficient fluctuations that the original Duan criterion~\cite{Duan2000a,Simon2000a} would not detect entanglement.  This effect becomes more prominent for larger numbers of particles (see the Appendix for further details). In this regime it is not possible to identify the signal state with  a simple two-mode squeezed state as generated by an optical parametric oscillator. The remarkable feature of these states is that the quadrature variances are smaller than theoretically possible using a coherent state local oscillator of the same size [c.f.~Eq.(\ref{moments})], a fact only made possible by the entanglement between local oscillator and signal beams (for further detail see the Appendix.)

Number fluctuations are an important consideration in an experimental setting, and will typically be at or above the shot-noise limit. We repeated the above simulation with an initial coherent state condensate with Poissonian statistics with a mean of $100$ atoms. We do not plot the results here as they are essentially identical to those in Fig.~\ref{fig:results}~(a--c) for an initial number state.  Simulations of systems beginning with a highly-mixed initial state with number fluctuations well above the shot-noise limit also demonstrate entanglement. We plot the results for an initial thermal (chaotic) distribution with a mean of 100 particles in Fig.~\ref{fig:results}~(d--f). We observe that in this case the maximal amount of violation and the range of values of $\chi t$ where entanglement is demonstrated are both reduced compared to an initial Fock state, but it is present nonetheless. In fact, for small values of $\chi t$ the violation slightly \emph{increases} with fluctuations. This result applies to both photons and atoms, and may suggest that the demonstration of entanglement in a range of quantum optics experiments are relatively unaffected by the intensity fluctuations of the laser sources.

In conclusion we have derived three new criteria for identifying continuous variable entanglement when local oscillators in arbitrary states are used in the quadrature measurements.  We have shown that these criteria can be violated and entanglement demonstrated for degenerate four-wave-mixing with as few as 100 particles.  In this situation the direct application of the criteria of Duan \emph{et al.}~\cite{Duan2000a}, Simon~\cite{Simon2000a} and Reid~\cite{Reid1989a} is inappropriate, and can lead to either falsely identifying entanglement, or failing to identify entanglement when the local oscillator modes are sufficiently small. We have also shown that initial number fluctuations will not necessarily prevent the demonstration of entanglement in experiment.

The authors would like to thank Margaret Reid, Ping Koy Lam and Claude Fabre for useful discussions, and acknowledge financial support from the Australian Research Council Centre of Excellence program and grant FF0458313.

\appendix

\section*{Appendix}

Here we expand on two issues discussed in the main text.
Firstly we compare and contrast the new entanglement criteria we derive to the earlier results of ~Duan \emph{et al.}~\cite{Duan2000a}, Simon~\cite{Simon2000a} and Reid~\cite{Reid1989a} when using \emph{coherent} state local oscillators of arbitrary size, and illustrate with an example.
Secondly, we present a more detailed analysis of degenerate four-wave mixing of a Bose-Einstein condensate as presented in the main text, and make a comparison for a range of particle numbers.

The issue of measuring quadrature variances using a limited size but uncorrelated local oscillator was addresed by Zhang \emph{et al.}~\cite{Zhang2001a}. In this experiment it was necessary to reduce the intensity of the local oscillator to be of a similar magnitude to the signal to avoid saturating the photodetector.  As is common in experiments in quantum optics, the local oscillator was separated from the beam generating the signal early in the experiment using a beam splitter. If the laser beam incident on the beam splitter is a coherent state (or can be written as a classical mixture of coherent states), then one can assume the signal and local oscillator beams are not correlated. We will examine the case where the source beam (and therefore local oscillator) is in a coherent state $|\beta\rangle$.

If the signal (local oscillator) mode has annihilation operator $\h{a}$  ($\h{b}$), the variance of the intensity difference of the homodyne measurement is
\begin{equation}
  \mathrm{Var}\bigl[\hd{a}\h{b} + \h{a}\hd{b}\bigr] = \langle \hd{b} \h{b} \rangle \mathrm{Var}\bigl[ \h{a} + \hd{a}
\bigr] + \langle \hd{a} \h{a} \rangle, \label{correction}
\end{equation}
where we have assumed $\langle \h{b} \rangle = \beta$ is real.  We can use this result to determine the canonical quadrature from the measured quadrature, and therefore derive the equivalent of the Duan \emph{et al.} inseparability criterion~\cite{Duan2000a} when using small coherent states for the local oscillators.
We find that  all separable states obey
\begin{multline} \label{Duan_coherent}
  \mathrm{Var}\bigl[\h{X}_1 \pm \h{X}_2\bigr] + \mathrm{Var}\bigl[\h{Y}_1 \mp \h{Y}_2\bigr]
  \ge \\ \;4 + 2\frac{\langle\hd{a}_1
\h{a}_1\rangle}{\langle\hd{b}_1\h{b}_1\rangle}
  + 2 \frac{\langle\hd{a}_2
\h{a}_2\rangle}{\langle\hd{b}_2\h{b}_2\rangle}.
\end{multline}
This inequality is easier to violate than the inseparability criterion we derive as Eq.~(7) in the main text, but requires that the local oscillators be in a coherent state.

The same corrections can be applied to derive an EPR criterion analogous to Eq.~(10) in the main text, assuming coherent state local oscillators
\begin{multline} \label{Duan_coherent2}
  \mathrm{Var}\bigl[\h{X}_1 \pm \h{X}_2\bigr] + \mathrm{Var}\bigl[\h{Y}_1 \mp \h{Y}_2\bigr]
  \ge \\ \;2 + 2\frac{\langle\hd{a}_1
\h{a}_1\rangle}{\langle\hd{b}_1\h{b}_1\rangle}
  + 2 \frac{\langle\hd{a}_2
\h{a}_2\rangle}{\langle\hd{b}_2\h{b}_2\rangle}.
\end{multline}
A similar analysis can be applied to the EPR criterion of Reid~\cite{Reid1989a}, with final result
\begin{gather}
\Delta^2_\mathrm{inf}\bigl[\h{X}_2\bigr] = \mathrm{Var}\bigl[ \h{X}_2 \bigr] - \frac{\langle\hd{a}_2
\h{a}_2\rangle}{\langle\hd{b}_2\h{b}_2\rangle}-
  \frac{\mathrm{Var}\bigl[\h{X}_1,\h{X}_2\bigr]^2}{\mathrm{Var}\bigl[\h{X}_1\bigr] - \frac{\langle\hd{a}_1
\h{a}_1\rangle}{\langle\hd{b}_1\h{b}_1\rangle}},\nonumber\\
  \Delta^2_\mathrm{inf}\bigl[\h{Y}_2\bigr] = \mathrm{Var}\bigl[ \h{Y}_2 \bigr] - \frac{\langle\hd{a}_2
\h{a}_2\rangle}{\langle\hd{b}_2\h{b}_2\rangle} - \frac{\mathrm{Var}\bigl[\h{Y}_1,\h{Y}_2\bigr]^2}{\mathrm{Var}\bigl[\h{Y}_1\bigr] - \frac{\langle\hd{a}_1
\h{a}_1\rangle}{\langle\hd{b}_1\h{b}_1\rangle}},\label{eq:EPR2} \\
    \Delta_{\mathrm{inf}}^2\bigl[\h{X}_2\bigr] \,
  \Delta_{\mathrm{inf}}^2\bigl[\h{Y}_2\bigr]
  \ge 1, \nonumber
\end{gather}
again assuming coherent state local oscillators.

The focus of this work is situations when it is \emph{not} possible to assume a coherent state local oscillator of \emph{any} size. The main text illustrated this with the example of degenerate four-wave mixing beginning from a number state of 100 bosons and using the remnants of this mode to form the local oscillators. ``Perfect''  local oscillators (i.e. large and coherent) for the quadrature measurements cannot be used to demonstrate entanglement in this system.  This can be seen from Eq.~(14) in the main text where the state basis clearly implies tripartite entanglement; using an external local oscillator is equivalent to tracing over mode 0 resulting in a mixed, separable density matrix.  Thus it is necessary to use mode 0 as the source of our local oscillators in order to detect entanglement in this system,
%This means the local oscillators are correlated (and in fact entangled) with the signal modes.
and the new entanglement criteria that we have derived can be correctly applied.

We now illustrate the important differences between using perfect local oscillators and the method of quadrature measurement described in the main text with a numerical simulation. To do this we begin with a coherent state in mode 0, and vacuum in modes 1 and 2, represented by the state-vector $|\alpha, 0, 0\rangle$. The state then evolves under Hamiltonian in Eq.~(13) of the main text, and so we can write the state vector in a reduced Fock basis,
\begin{equation}
  |\psi(t)\rangle = \sum_{N,m} c_{N,m}(t) |N-2m,m,m\rangle.
\end{equation}
The exact evolution can be computed efficiently as the Hamiltonian conserves total number and so the evolution of $\{ c_{N,m} \}$ can be calculated separately for each $N$. It is then a simple matter to extract the expectation values and variances of the quadratures and populations.

\begin{figure}
\begin{centering}
\includegraphics[width=6cm]{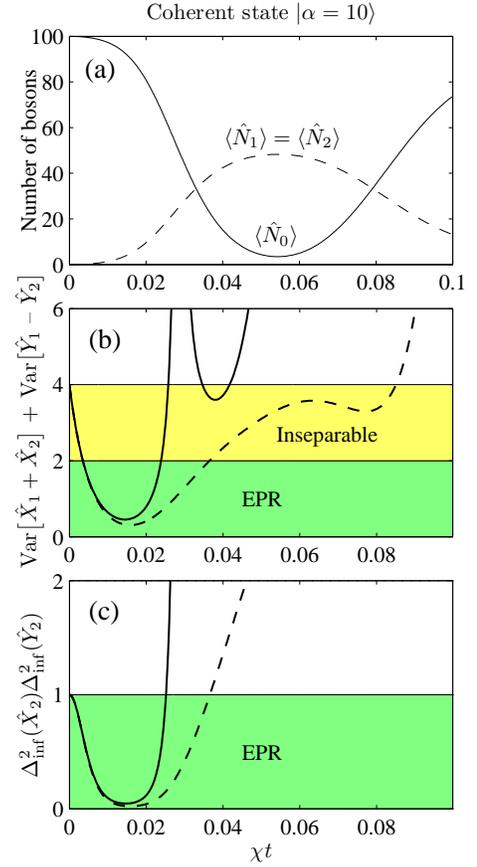}
\caption{(Color online) (a)~The population of the initial condensate (solid line) and the signal modes $1$ and $2$ (dashed line) as a function of time for degenerate four-wave mixing.  The condensate begins in a coherent state with a mean of 100 particles.
(b) The rescaled separability criterion [Eq.~(\ref{rescaled1})] is plotted (solid line) with a comparison with that of Duan \textit{et al.}~\cite{Duan2000a} (dashed line). (c)~The rescaled EPR criterion [Eq.~(\ref{rescaled3})] is plotted (solid line) with a comparison to the result of Reid~\cite{Reid1989a} (dashed line).} \label{fig:coherent}
\end{centering}
\end{figure}

\begin{figure}
\begin{centering}
\includegraphics[width=7cm]{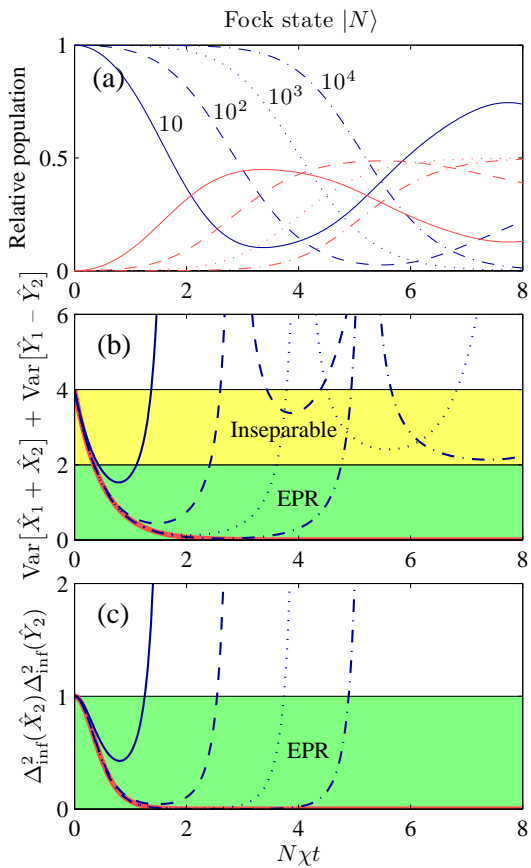}
\caption{(Color online) The population dynamics and entanglement criteria for Fock state initial conditions with 10 (solid line), 100 (dashed line), 1000 (dotted line) and 10000 (dash-dotted line) particles versus the scaled time $N\chi t$. (a)~In each case atoms are transferred from the condensate (blue lines) to modes $1$ and $2$ (red lines). (b)~The rescaled separability criterion [Eq.~(\ref{rescaled1}), blue lines] shows that the amount and duration (in scaled time units) of violation increases with particle number. Furthermore, the violation reoccurs after the local oscillator becomes smaller than modes 1 and 2. The undepleted pump solution is plotted for comparison (red line). (c)~The rescaled EPR criterion [Eq.~(\ref{rescaled3}), blue lines] shows similar behavior. The undepleted pump solution is also plotted (red line).} \label{fig:compare}
\end{centering}
\end{figure}

To make a direct comparison with the entanglement criteria of of Duan \emph{et al.}~\cite{Duan2000a}  we  rearrange our separability criterion (Eq.~(7) in the main text) to the following form
\begin{equation}
    \frac{\mathrm{Var}\bigl[\h{X}_1 \pm \h{X}_2\bigr] + \mathrm{Var}\bigl[\h{Y}_1 \mp \h{Y}_2\bigr]}{\frac{1}{2}\left| 1 - \frac{\langle\hd{a}_1
\h{a}_1\rangle}{\langle\hd{b}_1\h{b}_1\rangle} \right| + \frac{1}{2}\left| 1 - \frac{\langle\hd{a}_2
\h{a}_2\rangle}{\langle\hd{b}_2\h{b}_2\rangle} \right|}
  \ge 4, \label{rescaled1}
\end{equation}
For a comparison with the results of Reid~\cite{Reid1989a}, we rearrange the two EPR criteria (Eqs.~(10) and (8) in the main text)
\begin{gather}
    \frac{\mathrm{Var}\bigl[\h{X}_1 \pm \h{X}_2\bigr] + \mathrm{Var}\bigl[\h{Y}_1 \mp \h{Y}_2\bigr]}{\left| 1 - \frac{\langle\hd{a}_2
\h{a}_2\rangle}{\langle\hd{b}_2\h{b}_2\rangle} \right|} \ge 2, \label{rescaled2} \\
    \frac{\Delta_{\mathrm{inf}}^2\bigl[\h{X}_2\bigr] \,
  \Delta_{\mathrm{inf}}^2\bigl[\h{Y}_2\bigr]}{ \left( 1 - \frac{\langle \hd{a}_2 \h{a}_2
\rangle}{\langle\hd{b}_2\h{b}_2\rangle} \right)^2} \ge 1. \label{rescaled3}
\end{gather}
We note that $\langle \hd{a}_1 \h{a}_1
\rangle / \langle\hd{b}_1\h{b}_1\rangle = \langle \hd{a}_2 \h{a}_2
\rangle / \langle\hd{b}_2\h{b}_2\rangle$ for our system, and thus that the LHS of Eqs.~(\ref{rescaled1}) and (\ref{rescaled2}) are equal.
Setting the denominators on the LHS of Eqs.~(\ref{rescaled1},\ref{rescaled2},\ref{rescaled3})  to be one recovers the inequalities in Refs.~\cite{Duan2000a,Reid1989a}.
We plot a comparison of the different separability and EPR inequalities in Fig.~\ref{fig:coherent}.

Figure~\ref{fig:coherent}(a) shows the evolution of the mean population of the local oscillator and signal modes, which is identical for an initial condition of either a coherent state or a Poissonian mixture with a mean of 100 bosons. Figures~\ref{fig:coherent}(b,c) show a comparison between using the perfect local oscillator with previously known entanglement criteria \cite{Duan2000a,Reid1989a} and the corresponding results derived in this work. For short times they are in agreement, but later the perfect local oscillator clearly outperforms the remnants of mode 0. However, the point of maximum violation is not significantly different between the two methods.

We again note that for an initial Fock state, or any classical mixture of Fock states, the perfect local oscillator will be unable to detect any  entanglement between the signal modes. Also, if the state of an external local oscillator is somehow comprimised (e.g. by the Kerr nonlinearity, which is significant in BECs) then the remnants of mode 0 may make a more suitable phase reference for an experiment.

We now present a comparison of degenerate four-wave mixing with a BEC in an optical lattice for different initial particles numbers.
In Fig.~\ref{fig:compare} we plot a range of results from $N=10$ to $N=10000$ particles. For a clear comparison we have plotted the results on a time scale $N\chi t$ for which the \emph{absolute} populations of modes 1 and 2 are similar for all intial atom numbers at small times.  In this region the population of mode 0 is approximately constant with $\hat{a}_0^2\approx\hat{a}_0^{\dag 2} \approx N$, and so the Hamiltonian is
\begin{equation}
\h{H} \approx i\chi N \left(\h{a}_1\h{a}_2 - \hd{a}_1\hd{a}_2\right). \label{upa}
\end{equation}
This is the undepleted pump approximation with the exact solution in the Heisenberg picture
\begin{eqnarray}
  \h{a}_1(t) &=& \cosh(N\chi t)\h{a}_1(0) + \sinh(N\chi t)\hd{a}_2(0), \\
  \h{a}_2(t) &=& \cosh(N\chi t)\h{a}_2(0) + \sinh(N\chi t)\hd{a}_1(0).
\end{eqnarray}
This allows us to compare with the limit $N \rightarrow \infty$; we can also assume that mode 0 will act as a perfect local oscillator in this limit and apply the entanglement criteria of Duan \emph{et al.}~\cite{Duan2000a} and Reid~\cite{Reid1989a}.

Figure~\ref{fig:compare}(a) shows that the \emph{relative} number of atoms transferred from mode 0 to modes 1 and 2 remains insignificant for a longer timescale $N\chi t$ for larger initial populations. In Fig.~\ref{fig:compare}(b) and (c) we plot the rescaled entanglement criteria [Eqs.~(\ref{rescaled1},\ref{rescaled2},\ref{rescaled3})], and observe that the entanglement is detected for a larger range of $N\chi t$ for larger values of $N$. For comparison we have plotted the original entanglement criteria of Duan \emph{et al.}~\cite{Duan2000a} and Reid~\cite{Reid1989a} for the undepleted pump solution in red. The finite particle number solutions begin to deviate from this curve around the same time that the population of mode 0 decreases by a few percent.

In Fig.~\ref{fig:compare}(b) the separability criteria is violated in a region where modes 1 and 2 are larger than the local oscillator mode 0 for $N \ge 100$  particles. This violation increases with increased particle number, but we do not demonstrate the EPR paradox even with $N=10^4$ particles (corresponding to our computational limit). The state of the system at this time is complex and dissimilar to the two-mode squeezed state predicted by the undepleted pump approximation. It is interesting to note that at these times the system exhibits quadrature variances that are much smaller than is theoretically possible using a coherent local oscillator of the same size. We can see from Eq.~(\ref{correction}) that using a coherent state local oscillator sets a lower bound to the measured quadrature variance. This is also true for the sum or difference of quadratures
\begin{equation}
  \mathrm{Var}\bigl[\h{X}_1 \pm \h{X}_2\bigr] \ge \frac{\langle \hd{a}_1\h{a}_1\rangle}{\langle\hd{b}_1\h{b}_1\rangle} + \frac{\langle \hd{a}_2\h{a}_2\rangle}{\langle\hd{b}_2\h{b}_2\rangle}.
\end{equation}
The variances of the quadrature sums or differences predicted in our simulations are as small as half of this limit. The entanglement between the signal and local oscillator modes allows a reduced level of uncorrelated fluctuations.

Some time after the recurrence period the entanglement inequalities are again no longer violated once the population transfer from mode 0 to modes 1 and 2 has begun to reverse.  At this point the quantum state of the system has evolved quite significantly,  and we observe interference fringes in the number probability distribution for each of the modes.  These continue to become more complicated, and entanglement will not be observed at any later time.

%\bibliography{andy}

\begin{thebibliography}{34}
\expandafter\ifx\csname natexlab\endcsname\relax\def\natexlab#1{#1}\fi
\expandafter\ifx\csname bibnamefont\endcsname\relax
  \def\bibnamefont#1{#1}\fi
\expandafter\ifx\csname bibfnamefont\endcsname\relax
  \def\bibfnamefont#1{#1}\fi
\expandafter\ifx\csname citenamefont\endcsname\relax
  \def\citenamefont#1{#1}\fi
\expandafter\ifx\csname url\endcsname\relax
  \def\url#1{\texttt{#1}}\fi
\expandafter\ifx\csname urlprefix\endcsname\relax\def\urlprefix{URL }\fi
\providecommand{\bibinfo}[2]{#2}
\providecommand{\eprint}[2][]{\url{#2}}

\bibitem[{\citenamefont{Rolston and Phillips}(2002)}]{Rolston2002a}
\bibinfo{author}{\bibfnamefont{S.~L.} \bibnamefont{Rolston}} \bibnamefont{and}
  \bibinfo{author}{\bibfnamefont{W.~D.} \bibnamefont{Phillips}},
  \bibinfo{journal}{Nature} \textbf{\bibinfo{volume}{416}},
  \bibinfo{pages}{219} (\bibinfo{year}{2002}).

\bibitem[{\citenamefont{Molmer}(2003)}]{Molmer2003a}
\bibinfo{author}{\bibfnamefont{K.}~\bibnamefont{Molmer}}, \bibinfo{journal}{New
  J. Phys.} \textbf{\bibinfo{volume}{5}}, \bibinfo{pages}{55}
  (\bibinfo{year}{2003}).

\bibitem[{\citenamefont{Schellekens et~al.}(2005)\citenamefont{Schellekens,
  Hoppeler, Perrin, Gomes, Boiron, Aspect, and Westbrook}}]{Schellekens2005a}
\bibinfo{author}{\bibfnamefont{M.}~\bibnamefont{Schellekens}},
  \bibinfo{author}{\bibfnamefont{R.}~\bibnamefont{Hoppeler}},
  \bibinfo{author}{\bibfnamefont{A.}~\bibnamefont{Perrin}},
  \bibinfo{author}{\bibfnamefont{J.~V.} \bibnamefont{Gomes}},
  \bibinfo{author}{\bibfnamefont{D.}~\bibnamefont{Boiron}},
  \bibinfo{author}{\bibfnamefont{A.}~\bibnamefont{Aspect}}, \bibnamefont{and}
  \bibinfo{author}{\bibfnamefont{C.~I.} \bibnamefont{Westbrook}},
  \bibinfo{journal}{Science} \textbf{\bibinfo{volume}{310}},
  \bibinfo{pages}{648} (\bibinfo{year}{2005}).

\bibitem[{\citenamefont{Jeltes et~al.}(2007)\citenamefont{Jeltes, McNamara,
  Hogervorst, Vassen, Krachmalnicoff, Schellekens, Perrin, Chang, Boiron,
  Aspect et~al.}}]{Vassen2007a}
\bibinfo{author}{\bibfnamefont{T.}~\bibnamefont{Jeltes}},
  \bibinfo{author}{\bibfnamefont{J.~M.} \bibnamefont{McNamara}},
  \bibinfo{author}{\bibfnamefont{W.}~\bibnamefont{Hogervorst}},
  \bibinfo{author}{\bibfnamefont{W.}~\bibnamefont{Vassen}},
  \bibinfo{author}{\bibfnamefont{V.}~\bibnamefont{Krachmalnicoff}},
  \bibinfo{author}{\bibfnamefont{M.}~\bibnamefont{Schellekens}},
  \bibinfo{author}{\bibfnamefont{A.}~\bibnamefont{Perrin}},
  \bibinfo{author}{\bibfnamefont{H.}~\bibnamefont{Chang}},
  \bibinfo{author}{\bibfnamefont{D.}~\bibnamefont{Boiron}},
  \bibinfo{author}{\bibfnamefont{A.}~\bibnamefont{Aspect}},
  \bibnamefont{et~al.}, \bibinfo{journal}{Nature (London)}
  \textbf{\bibinfo{volume}{445}}, \bibinfo{pages}{402} (\bibinfo{year}{2007}).

\bibitem[{\citenamefont{Chuu et~al.}(2005)\citenamefont{Chuu, Schreck, Meyrath,
  Hanssen, Price, and Raizen}}]{Chuu2005a}
\bibinfo{author}{\bibfnamefont{C.-S.} \bibnamefont{Chuu}},
  \bibinfo{author}{\bibfnamefont{F.}~\bibnamefont{Schreck}},
  \bibinfo{author}{\bibfnamefont{T.~P.} \bibnamefont{Meyrath}},
  \bibinfo{author}{\bibfnamefont{J.~L.} \bibnamefont{Hanssen}},
  \bibinfo{author}{\bibfnamefont{G.~N.} \bibnamefont{Price}}, \bibnamefont{and}
  \bibinfo{author}{\bibfnamefont{M.~G.} \bibnamefont{Raizen}},
  \bibinfo{journal}{Phys. Rev. Lett.} \textbf{\bibinfo{volume}{95}},
  \bibinfo{pages}{260403} (\bibinfo{year}{2005}).

\bibitem[{\citenamefont{Greiner et~al.}(2005)\citenamefont{Greiner, Regal,
  Stewart, and Jin}}]{Greiner2005a}
\bibinfo{author}{\bibfnamefont{M.}~\bibnamefont{Greiner}},
  \bibinfo{author}{\bibfnamefont{C.~A.} \bibnamefont{Regal}},
  \bibinfo{author}{\bibfnamefont{J.~T.} \bibnamefont{Stewart}},
  \bibnamefont{and} \bibinfo{author}{\bibfnamefont{D.~S.} \bibnamefont{Jin}},
  \bibinfo{journal}{Phys. Rev. Lett.} \textbf{\bibinfo{volume}{94}},
  \bibinfo{pages}{110401} (\bibinfo{year}{2005}).

\bibitem[{\citenamefont{F{\"o}lling et~al.}(2005)\citenamefont{F{\"o}lling,
  Gerbier, Widera, Mandel, Gericke, and Bloch}}]{Folling2005a}
\bibinfo{author}{\bibfnamefont{S.}~\bibnamefont{F{\"o}lling}},
  \bibinfo{author}{\bibfnamefont{F.}~\bibnamefont{Gerbier}},
  \bibinfo{author}{\bibfnamefont{A.}~\bibnamefont{Widera}},
  \bibinfo{author}{\bibfnamefont{O.}~\bibnamefont{Mandel}},
  \bibinfo{author}{\bibfnamefont{T.}~\bibnamefont{Gericke}}, \bibnamefont{and}
  \bibinfo{author}{\bibfnamefont{I.}~\bibnamefont{Bloch}},
  \bibinfo{journal}{Nature (London)} \textbf{\bibinfo{volume}{434}},
  \bibinfo{pages}{481} (\bibinfo{year}{2005}).

\bibitem[{\citenamefont{Schr{\"o}dinger}(1935)}]{Schroedinger1935a}
\bibinfo{author}{\bibfnamefont{E.}~\bibnamefont{Schr{\"o}dinger}},
  \bibinfo{journal}{Proceedings of the Cambridge Philosophical Society}
  \textbf{\bibinfo{volume}{31}}, \bibinfo{pages}{555} (\bibinfo{year}{1935}).

\bibitem[{\citenamefont{Duan et~al.}(2000)\citenamefont{Duan, Giedke, Cirac,
  and Zoller}}]{Duan2000a}
\bibinfo{author}{\bibfnamefont{L.~M.} \bibnamefont{Duan}},
  \bibinfo{author}{\bibfnamefont{G.}~\bibnamefont{Giedke}},
  \bibinfo{author}{\bibfnamefont{J.~I.} \bibnamefont{Cirac}}, \bibnamefont{and}
  \bibinfo{author}{\bibfnamefont{P.}~\bibnamefont{Zoller}},
  \bibinfo{journal}{Phys. Rev. Lett.} \textbf{\bibinfo{volume}{84}},
  \bibinfo{pages}{2722} (\bibinfo{year}{2000}).

\bibitem[{\citenamefont{Simon}(2000)}]{Simon2000a}
\bibinfo{author}{\bibfnamefont{R.}~\bibnamefont{Simon}},
  \bibinfo{journal}{Phys. Rev. Lett.} \textbf{\bibinfo{volume}{84}},
  \bibinfo{pages}{2726} (\bibinfo{year}{2000}).

\bibitem[{\citenamefont{Reid}(1989)}]{Reid1989a}
\bibinfo{author}{\bibfnamefont{M.~D.} \bibnamefont{Reid}},
  \bibinfo{journal}{Phys. Rev. A} \textbf{\bibinfo{volume}{40}},
  \bibinfo{pages}{913} (\bibinfo{year}{1989}).

\bibitem[{\citenamefont{Einstein et~al.}(1935)\citenamefont{Einstein, Podolsky,
  and Rosen}}]{Einstein1935a}
\bibinfo{author}{\bibfnamefont{A.}~\bibnamefont{Einstein}},
  \bibinfo{author}{\bibfnamefont{B.}~\bibnamefont{Podolsky}}, \bibnamefont{and}
  \bibinfo{author}{\bibfnamefont{N.}~\bibnamefont{Rosen}},
  \bibinfo{journal}{Phys. Rev.} \textbf{\bibinfo{volume}{47}},
  \bibinfo{pages}{777} (\bibinfo{year}{1935}).

\bibitem[{\citenamefont{Gardiner and Zoller}(2004)}]{QuantumNoise}
\bibinfo{author}{\bibfnamefont{C.~W.} \bibnamefont{Gardiner}} \bibnamefont{and}
  \bibinfo{author}{\bibfnamefont{P.}~\bibnamefont{Zoller}},
  \emph{\bibinfo{title}{Quantum Noise}} (\bibinfo{publisher}{Springer},
  \bibinfo{address}{Berlin}, \bibinfo{year}{2004}), \bibinfo{edition}{3rd} ed.

\bibitem[{\citenamefont{Slusher et~al.}(1985)\citenamefont{Slusher, Hollberg,
  Yurke, Mertz, and Valley}}]{Shlusher1985a}
\bibinfo{author}{\bibfnamefont{R.~E.} \bibnamefont{Slusher}},
  \bibinfo{author}{\bibfnamefont{L.~W.} \bibnamefont{Hollberg}},
  \bibinfo{author}{\bibfnamefont{B.}~\bibnamefont{Yurke}},
  \bibinfo{author}{\bibfnamefont{J.~C.} \bibnamefont{Mertz}}, \bibnamefont{and}
  \bibinfo{author}{\bibfnamefont{J.~F.} \bibnamefont{Valley}},
  \bibinfo{journal}{Phys. Rev. Lett.} \textbf{\bibinfo{volume}{55}},
  \bibinfo{pages}{2409} (\bibinfo{year}{1985}).

\bibitem[{\citenamefont{Zhang et~al.}(2001)\citenamefont{Zhang, Coudrea,
  Martinelli, Ma\^{i}tre, and Fabre}}]{Zhang2001a}
\bibinfo{author}{\bibfnamefont{K.~S.} \bibnamefont{Zhang}},
  \bibinfo{author}{\bibfnamefont{T.}~\bibnamefont{Coudrea}},
  \bibinfo{author}{\bibfnamefont{M.}~\bibnamefont{Martinelli}},
  \bibinfo{author}{\bibfnamefont{A.}~\bibnamefont{Ma\^{i}tre}},
  \bibnamefont{and} \bibinfo{author}{\bibfnamefont{C.}~\bibnamefont{Fabre}},
  \bibinfo{journal}{Phys. Rev. Lett.} \textbf{\bibinfo{volume}{64}},
  \bibinfo{pages}{033815} (\bibinfo{year}{2001}).

\bibitem[{\citenamefont{Ou et~al.}(1992)\citenamefont{Ou, Pereira, Kimble, and
  Peng}}]{Ou1992a}
\bibinfo{author}{\bibfnamefont{Z.~Y.} \bibnamefont{Ou}},
  \bibinfo{author}{\bibfnamefont{S.~F.} \bibnamefont{Pereira}},
  \bibinfo{author}{\bibfnamefont{H.~J.} \bibnamefont{Kimble}},
  \bibnamefont{and} \bibinfo{author}{\bibfnamefont{K.~C.} \bibnamefont{Peng}},
  \bibinfo{journal}{Phys. Rev. Lett.} \textbf{\bibinfo{volume}{68}},
  \bibinfo{pages}{3663} (\bibinfo{year}{1992}).

\bibitem[{\citenamefont{Smithey et~al.}(1993)\citenamefont{Smithey, Beck,
  Raymer, and Faridani}}]{Smithey1993a}
\bibinfo{author}{\bibfnamefont{D.~T.} \bibnamefont{Smithey}},
  \bibinfo{author}{\bibfnamefont{M.}~\bibnamefont{Beck}},
  \bibinfo{author}{\bibfnamefont{M.~G.} \bibnamefont{Raymer}},
  \bibnamefont{and} \bibinfo{author}{\bibfnamefont{A.}~\bibnamefont{Faridani}},
  \bibinfo{journal}{Phys. Rev. Lett.} \textbf{\bibinfo{volume}{70}},
  \bibinfo{pages}{1244} (\bibinfo{year}{1993}).

\bibitem[{\citenamefont{Bowen et~al.}(2003)\citenamefont{Bowen, Treps, Buchler,
  Schnabel, Ralph, Bachor, Symul, and Lam}}]{Bowen2003a}
\bibinfo{author}{\bibfnamefont{W.~P.} \bibnamefont{Bowen}},
  \bibinfo{author}{\bibfnamefont{N.}~\bibnamefont{Treps}},
  \bibinfo{author}{\bibfnamefont{B.~C.} \bibnamefont{Buchler}},
  \bibinfo{author}{\bibfnamefont{R.}~\bibnamefont{Schnabel}},
  \bibinfo{author}{\bibfnamefont{T.~C.} \bibnamefont{Ralph}},
  \bibinfo{author}{\bibfnamefont{H.-A.} \bibnamefont{Bachor}},
  \bibinfo{author}{\bibfnamefont{T.}~\bibnamefont{Symul}}, \bibnamefont{and}
  \bibinfo{author}{\bibfnamefont{P.~K.} \bibnamefont{Lam}},
  \bibinfo{journal}{Phys. Rev. A} \textbf{\bibinfo{volume}{67}},
  \bibinfo{pages}{032302} (\bibinfo{year}{2003}).

\bibitem[{\citenamefont{Hilligs{\o}e and M{\o}lmer}(2005)}]{Hilligsoe2005a}
\bibinfo{author}{\bibfnamefont{K.~M.} \bibnamefont{Hilligs{\o}e}}
  \bibnamefont{and}
  \bibinfo{author}{\bibfnamefont{K.}~\bibnamefont{M{\o}lmer}},
  \bibinfo{journal}{Phys. Rev. A} \textbf{\bibinfo{volume}{71}},
  \bibinfo{pages}{041602(R)} (\bibinfo{year}{2005}).

\bibitem[{\citenamefont{Campbell et~al.}(2006)\citenamefont{Campbell, Mun,
  Boyd, Streed, Ketterle, and Pritchard}}]{Campbell2006a}
\bibinfo{author}{\bibfnamefont{G.}~\bibnamefont{Campbell}},
  \bibinfo{author}{\bibfnamefont{J.}~\bibnamefont{Mun}},
  \bibinfo{author}{\bibfnamefont{M.}~\bibnamefont{Boyd}},
  \bibinfo{author}{\bibfnamefont{E.}~\bibnamefont{Streed}},
  \bibinfo{author}{\bibfnamefont{W.}~\bibnamefont{Ketterle}}, \bibnamefont{and}
  \bibinfo{author}{\bibfnamefont{D.}~\bibnamefont{Pritchard}},
  \bibinfo{journal}{Phys. Rev. Lett.} \textbf{\bibinfo{volume}{96}},
  \bibinfo{pages}{020406} (\bibinfo{year}{2006}).

\bibitem[{\citenamefont{Olsen and Davis}(2006)}]{Olsen2006a}
\bibinfo{author}{\bibfnamefont{M.~K.} \bibnamefont{Olsen}} \bibnamefont{and}
  \bibinfo{author}{\bibfnamefont{M.~J.} \bibnamefont{Davis}},
  \bibinfo{journal}{Phys. Rev. A} \textbf{\bibinfo{volume}{73}},
  \bibinfo{pages}{063618} (\bibinfo{year}{2006}).

\bibitem[{\citenamefont{Kheruntsyan et~al.}(2005)\citenamefont{Kheruntsyan,
  Olsen, and Drummond}}]{Kheruntsyan2005a}
\bibinfo{author}{\bibfnamefont{K.~V.} \bibnamefont{Kheruntsyan}},
  \bibinfo{author}{\bibfnamefont{M.~K.} \bibnamefont{Olsen}}, \bibnamefont{and}
  \bibinfo{author}{\bibfnamefont{P.~D.} \bibnamefont{Drummond}},
  \bibinfo{journal}{Phys. Rev. Lett.} \textbf{\bibinfo{volume}{95}},
  \bibinfo{pages}{150405} (\bibinfo{year}{2005}).

\bibitem[{\citenamefont{Haine and Hope}(2005)}]{Haine2005a}
\bibinfo{author}{\bibfnamefont{S.~A.} \bibnamefont{Haine}} \bibnamefont{and}
  \bibinfo{author}{\bibfnamefont{J.~J.} \bibnamefont{Hope}},
  \bibinfo{journal}{Phys. Rev. A} \textbf{\bibinfo{volume}{72}},
  \bibinfo{pages}{033601} (\bibinfo{year}{2005}).

\bibitem[{\citenamefont{Braunstein and van Loock}(2005)}]{Braunstein2005a}
\bibinfo{author}{\bibfnamefont{S.~L.} \bibnamefont{Braunstein}}
  \bibnamefont{and} \bibinfo{author}{\bibfnamefont{P.}~\bibnamefont{van
  Loock}}, \bibinfo{journal}{Rev. Mod. Phys.} \textbf{\bibinfo{volume}{77}},
  \bibinfo{pages}{513} (\bibinfo{year}{2005}).

\bibitem[{\citenamefont{de~Oliveira}(2003)}]{deOliveira2003a}
\bibinfo{author}{\bibfnamefont{M.~C.} \bibnamefont{de~Oliveira}},
  \bibinfo{journal}{Phys. Rev. A} \textbf{\bibinfo{volume}{67}},
  \bibinfo{pages}{022307} (\bibinfo{year}{2003}).

\bibitem[{\citenamefont{da~Cunha and de~Oliveira}(2007)}]{daCunha2007a}
\bibinfo{author}{\bibfnamefont{B.~R.} \bibnamefont{da~Cunha}} \bibnamefont{and}
  \bibinfo{author}{\bibfnamefont{M.~C.} \bibnamefont{de~Oliveira}},
  \bibinfo{journal}{Phys. Rev. A} \textbf{\bibinfo{volume}{75}},
  \bibinfo{pages}{063615} (\bibinfo{year}{2007}).

\bibitem[{\citenamefont{van~der Stam et~al.}(2007)\citenamefont{van~der Stam,
  Meppelink, Vogels, and van~der Straten}}]{vanDerStam2007a}
\bibinfo{author}{\bibfnamefont{K.~M.~R.} \bibnamefont{van~der Stam}},
  \bibinfo{author}{\bibfnamefont{R.}~\bibnamefont{Meppelink}},
  \bibinfo{author}{\bibfnamefont{J.~M.} \bibnamefont{Vogels}},
  \bibnamefont{and} \bibinfo{author}{\bibfnamefont{P.}~\bibnamefont{van~der
  Straten}}, \bibinfo{journal}{Phys. Rev. A} \textbf{\bibinfo{volume}{75}},
  \bibinfo{pages}{031602} (\bibinfo{year}{2007}).

\bibitem[{\citenamefont{Hofmann and Takeuchi}(2003)}]{Hofmann2003a}
\bibinfo{author}{\bibfnamefont{H.~F.} \bibnamefont{Hofmann}} \bibnamefont{and}
  \bibinfo{author}{\bibfnamefont{S.}~\bibnamefont{Takeuchi}},
  \bibinfo{journal}{Phys. Rev. A} \textbf{\bibinfo{volume}{68}},
  \bibinfo{pages}{032103} (\bibinfo{year}{2003}).

\bibitem[{\citenamefont{Raymer et~al.}(2003)\citenamefont{Raymer, Funk,
  Sanders, and de~Guise}}]{Raymer2003a}
\bibinfo{author}{\bibfnamefont{M.~G.} \bibnamefont{Raymer}},
  \bibinfo{author}{\bibfnamefont{A.~C.} \bibnamefont{Funk}},
  \bibinfo{author}{\bibfnamefont{B.~C.} \bibnamefont{Sanders}},
  \bibnamefont{and} \bibinfo{author}{\bibfnamefont{H.}~\bibnamefont{de~Guise}},
  \bibinfo{journal}{Phys. Rev. A} \textbf{\bibinfo{volume}{67}},
  \bibinfo{pages}{052104} (\bibinfo{year}{2003}).

%\bibitem[{EPA()}]{EPAPS_Ferris2008b}
%\bibinfo{note}{See EPAPS Document No. ?? for a coherent state local oscillator
%  comparison and for additional numerical results.}

\bibitem[{\citenamefont{Wiseman et~al.}(2007)\citenamefont{Wiseman, Jones, and
  Doherty}}]{Wiseman2007a}
\bibinfo{author}{\bibfnamefont{H.~M.} \bibnamefont{Wiseman}},
  \bibinfo{author}{\bibfnamefont{S.~J.} \bibnamefont{Jones}}, \bibnamefont{and}
  \bibinfo{author}{\bibfnamefont{A.~C.} \bibnamefont{Doherty}},
  \bibinfo{journal}{Phys. Rev. Lett.} \textbf{\bibinfo{volume}{98}},
  \bibinfo{pages}{140402} (\bibinfo{year}{2007}).

\bibitem[{\citenamefont{Jones et~al.}(2007)\citenamefont{Jones, Wiseman, and
  Doherty}}]{Jones2007a}
\bibinfo{author}{\bibfnamefont{S.~J.} \bibnamefont{Jones}},
  \bibinfo{author}{\bibfnamefont{H.~M.} \bibnamefont{Wiseman}},
  \bibnamefont{and} \bibinfo{author}{\bibfnamefont{A.~C.}
  \bibnamefont{Doherty}}, \bibinfo{journal}{Phys. Rev. A}
  \textbf{\bibinfo{volume}{76}}, \bibinfo{pages}{052116}
  (\bibinfo{year}{2007}).

\bibitem[{\citenamefont{Reid et~al.}(2008)\citenamefont{Reid, Drummond,
  Cavalcanti, Bowen, Lam, Bachor, Anderson, and Leuchs}}]{ReidReview2008}
\bibinfo{author}{\bibfnamefont{M.~D.} \bibnamefont{Reid}},
  \bibinfo{author}{\bibfnamefont{P.~D.} \bibnamefont{Drummond}},
  \bibinfo{author}{\bibfnamefont{E.~G.} \bibnamefont{Cavalcanti}},
  \bibinfo{author}{\bibfnamefont{W.~P.} \bibnamefont{Bowen}},
  \bibinfo{author}{\bibfnamefont{P.~K.} \bibnamefont{Lam}},
  \bibinfo{author}{\bibfnamefont{H.~A.} \bibnamefont{Bachor}},
  \bibinfo{author}{\bibfnamefont{U.~L.} \bibnamefont{Anderson}},
  \bibnamefont{and} \bibinfo{author}{\bibfnamefont{G.}~\bibnamefont{Leuchs}},
  \bibinfo{journal}{eprint arXiv:0806.0270v1}  (\bibinfo{year}{2008}).

\bibitem[{\citenamefont{Cavalcanti et~al.}(2008)\citenamefont{Cavalcanti,
  Jones, Wiseman, and Reid}}]{CavalcantiUnpublished2008}
\bibinfo{author}{\bibfnamefont{E.~G.} \bibnamefont{Cavalcanti}},
  \bibinfo{author}{\bibfnamefont{S.~J.} \bibnamefont{Jones}},
  \bibinfo{author}{\bibfnamefont{H.~M.} \bibnamefont{Wiseman}},
  \bibnamefont{and} \bibinfo{author}{\bibfnamefont{M.~D.} \bibnamefont{Reid}},
  \bibinfo{journal}{in preparation}  (\bibinfo{year}{2008}).

\end{thebibliography}
%\bibliographystyle{apsrev}

% We are allowed shorten the references according to the Style and Notation guide.
% Only if length constrained and four or more authors:
% J. M. Smith \emph{et al.}, Phys. Rev. B \bf{46}, 1 (1992).

\end{document}